\documentstyle[prb,epsfig,aps]{revtex}
\begin{document}
\draft
\title{From pseudomorphic to orthomorphic growth of Fe films on Cu$_3$Au(001).}

\author{
F. Bruno$^{a}$, S. Terreni$^{b}$, L. Floreano$^{a}$, A. 
Cossaro,$^{a}$ D. Cvetko$^{a,c}$, 
P. Luches$^{d}$, L. Mattera$^{b}$, A. Morgante$^{a,e}$, R. Moroni$^{b}$, 
M. Repetto,$^{b}$ A. Verdini$^{a}$, and M. Canepa$^{b,*}$ }

\address{
$^{a}$Laboratorio TASC dell'Istituto Nazionale per la Fisica della
Materia, Trieste, Italy.\\
$^{b}$INFM and Dipartimento di Fisica, Universit\`a di Genova,\\
 via Dodecaneso 33, I-16146 Genova, Italy\\
$^{c}$J. Stefan Institute,
University of Ljubljana, Slovenia, and Sincrotrone Trieste,Italy\\
$^{d}$ INFM and Dipartimento di Fisica, Universit\`a di Modena e Reggio Emilia,
 Modena, Italy.\\
$^{e}$ Dipartimento di Fisica dell'Universit\`a di Trieste, Italy. \\
}

\maketitle

\begin{abstract}

The structure of Fe films grown on the (001) surface of a ${\rm Cu_{3}Au}$
single crystal at room temperature has been investigated by means of
Grazing Incidence X-Ray Diffraction (GIXRD) and Photo/Auger--Electron
Diffraction (ED) as a function of thickness in the 3-36~\AA\ range. The combination
of GIXRD and ED allows one to obtain quantitative information on the in--plane
spacing $a$ from the former technique, and the ratio between the vertical
spacing $c$ and $a$, from the latter one. At low coverage the film grows
 pseudomorphic to the face centered cubic substrate.
The experimental results obtained on a film of 8~\AA~ thickness clearly 
indicate the overcoming of the limit for pseudomorphic growth.
Above this limit the film is characterized by the coexistence of the
 pseudomorphic phase with another tetragonally strained phase $\gamma$, which 
 falls on the epitaxial line of ferromagnetic face-centered cubic Fe. 
Finally,  the development of a  body-centered 
phase $\alpha$, whose unit cell is rotated of 45$^{\circ }$ with
respect to the substrate one, has been clearly 
observed at $\sim$ 17 ~\AA. $\alpha$  is the 
dominating phase for  film thickness above $\sim$ 25 ~\AA~   
and its lattice constant evolves towards the orthomorphic phase in 
strict quantitative agreement with  
epitaxial curves calculated for body centered tetragonal 
iron phases.
\\
\\

\end{abstract}

$^{*}¥$ Corresponding author: Tel: **39-010-3536287; Fax: 
**39-010-311066;\\

\textit{E-mail address:} canepa@fisica.unige.it; wwww.fisica.unige.it/surfmag\\

\twocolumn
\narrowtext

\section*{Introduction}

Very thin iron films, with physical properties at variance with the ordinary 
$\alpha$--Fe phase, can be grown on appropriately chosen substrates.\cite
{wufeflo} Such ''artificial'' phases represent an ideal ground in the
investigation of the interplay between the structure and the magnetic
properties of materials. \cite{Moruzzi} In this field, the Cu$_3$Au(001)
 surface has been considered as a candidate for the stabilisation of a
face--centered--cubic (fcc), expanded volume, high--spin Fe phase.\cite
{earlyjona} For this substrate, a model of the so--called epitaxial lines of
iron, calculated within the frame of linear elasticity theory, suggested the
formation of a tetragonally strained phase, with some uncertainty on its
body or face centered geometry.\cite{jonamarcus} The calculations did not
consider interdiffusion processes between the Fe deposit and surface
atoms, but both intermixing at the interface and substrate segregation have
been reported for Fe deposited at room temperature (RT) on Cu(001)\cite
{fecuintermix} and on Au(001).\cite{opitz,feau}

In previous research on the magnetism of the Fe/Cu$_3$Au(001) films,
three regions of thickness of different properties were identified.\cite
{kisker1,kirschnerM,wuttigM} At sub--monolayer coverage no hysteresis loops
were detected. A second region was instead characterised by a magnetisation
perpendicular to the surface. Then, at a critical thickness $\Theta _{sw}$
of the order of a few monolayers and depending on the temperature of
deposition, the magnetisation was found to switch to an in plane orientation
( spin reorientation transition, in brief SRO) as in ordinary $\alpha $--Fe.

In spite of an overall agreement on film magnetism, no general consensus has
been reached about the film structure. Regarding room temperature (RT)
growth, an early LEED study claimed the occurrence of a face--centered cubic
phase up to 7 Mono--Layers (ML).\cite{kisker1} In a more recent LEED I-V
investigation the SRO transition ($\Theta _{sw}$ $\sim 4$ ML) was tightly
related to a transition from an fcc to a bcc--like phase;\cite{kirschnerM}
the LEED analysis was successively backed by an STM study in the coverage
range of the SRO transition, pointing out a complex topography assigned to
the coexistence of different phases.\cite{Lin} Another LEED I-V study,
backed by dynamical calculations, proposed instead a
body--centered--tetragonal (bct) structure down to $\Theta $ = 3.3~ML with
no apparent correlation between structural properties and the SRO transition.
\cite{wuttigS} In fact, the latter experiments must be carefully 
taken into consideration, since they were performed on substrates 
prepared by pre-depositing 2~ML of Fe at 150~K. For this temperature 
and thickness range, we have recently shown that the structure and 
morphology of of the growing film is driven by electronic 
mechanisms,\cite{canepa} 
whereas only strain and thermally activated processes drive the RT 
growth.
Finally, in a recent electron scattering experiment (in the
so--called Primary---beam Diffraction Modulated Electron Emission
configuration, PDMEE) \cite{luches} an fcc--like structure was 
found for $\Theta$ $<$ ~4~ML; 
then in a region extending up to $\approx $ 23~ML, two
different phases were detected. A unique phase of bct type was finally
detected, at least within the outermost layers, for relatively thicker
films, in agreement with results of Ref.\cite{wuttigS}. Ref.\cite{luches}
considered atomic exchange processes in some details, indicating the
occurrence of a limited Au intermixing and segregation  at the initial stages
of growth.

The somewhat conflicting results demand for further investigation. Here we
report on experiments of Fe/${\rm Cu_3Au(001)}$ RT\ growth performed at the
ALOISA beamline at the Elettra Synchrotron (Trieste, Italy).\cite{aloisaweb}
Our investigation deals with a structural characterization of films in a
thickness range of 3-36~\AA , performed by means of X-ray induced photo--
(and/or Auger--) electron Diffraction (from now on ED for brevity) in the
so--called forward scattering condition. In this condition, the emission
intensity is enhanced along the direction of interatomic axes by means of a
focusing effect.\cite{focusing} The angular position of focusing peaks in
polar ED scans is well known to provide ''simple'', chemically selected,
short range information on the structure of the topmost layers of the films.
\cite{revped1,revped2} Furthermore, the availability of reliable
calculations codes allows for a close comparison with experimental data and
for quantitative analysis.\cite{codes}

In our experimental approach, \cite{ass} ED data are backed by Grazing
Incidence X-Rays Diffraction (GIXRD) measurements, supplying information
about the in--plane lattice parameter of films \cite{inplanexrd} and about
the film morphology. In the next section details on the experimental
procedures are reported. Experimental data will be described in sect. \ref
{sec:risultati}. A quantitative analysis of the data can be found in the
sect. \ref{sec:analisidati}. A discussion of the results obtained follows in
sect. \ref{sec:discussione}. A summary and the conclusions can be found in
sect. \ref{sec:conclusioni}.

\section{Experimental}

The beamline ALOISA allows users to perform both electron spectroscopy and
X-Ray surface diffraction measurements under the same experimental
conditions.\cite{aloisaweb,aloisa} The sample is mounted on a
6--degrees--of--freedom manipulator, specially designed to select with great
accuracy (0.01$^{\circ }$) the grazing angle of the beam electric field. The
temperature of the sample, measured by thermocouples, can be varied by
resistive heating and liquid nitrogen cooling. The UHV experimental chamber
(base pressure in the 10$^{-11}$~mBar range) hosts the hemispherical
electron analysers and X-Ray detectors. The emission direction from the
sample surface can be freely selected for any orientation of the surface.
For the surface preparation, the sample is translated in the preparation
chamber (base pressure of $1-2 \times 10^{-10}$~mBar) equipped with
facilities for sputtering, evaporation cells, gas inlets and a RHEED system.

The surface preparation procedure was set up in previous He diffraction
experiments.\cite{scimia,eurolett} The procedure takes into account the
particular thermodynamics of ${\rm Cu_3Au(001)}$, that is characterized by a
continuous order/disorder (O/D) phase transition at the surface with a
critical temperature T$_c$ = 663 K \cite{eurolett} and a bulk first order
O/D transition at the same T$_c$.\cite{dosch} An ordered surface, displaying
sharp $c(2\times 2)$ RHEED patterns typical of the Au-Cu
termination,\cite{macrae} was obtained by sputtering and careful annealing
procedures described in details elsewhere.\cite{eurolett} The same sample
was also used in a previous synchrotron experiment \cite{ass} and in the
PDMEE experiment of Ref. \cite{luches}.

At ALOISA, XPS surveys at grazing incidence (of the order of the critical
angle) were used to check contamination of light adsorbates and Fe residuals
after the sputtering removal of films. A preliminary ED characterization of
the substrate, backed by multiple scattering calculations has been presented
in a previous paper.\cite{ass} Simulations taking into account the geometry
for the Au-Cu termination available in literature \cite{cuaustru} were found
to be in excellent agreement with experimental results.

\begin{figure}[tbp]
\includegraphics[width=.5\textwidth]{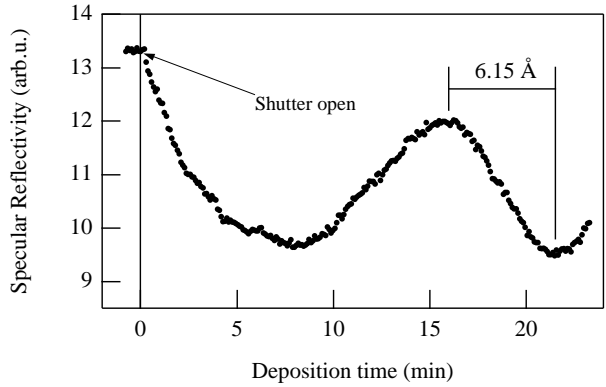}
\caption{X-Ray Specular Reflectivity taken during deposition at fixed 
energy (3500 eV) and grazing angle (8.25$^{\circ}$). Maxima and minima 
arise from the interference between the interface and the growing film 
surface.}
\label{fig1}
\end{figure}

Iron was evaporated from a carefully outgassed electron bombardment cell
(Omicron). A quartz microbalance allowed to tune the deposition flux
(typically of the order of 1 layer per minute) prior to the deposition on
the sample. The quartz microbalance was calibrated by X-ray reflectivity 
from the sample during and after deposition. 
Fig.~\ref{fig1} shows a typical deposition curve, measured
at an incidence (tilt) angle $\alpha _{in}$~=~8.25$^{\circ}$ and a
photon energy of 3500 eV during film deposition at room temperature. The
interference between the waves reflected by the film-vacuum and the
substrate-film interfaces yields an oscillatory evolution of the
reflectivity as a function of the overlayer thickness. For the selected
vertical momentum transfer $k_z=2\frac E{\hbar c}\sin \alpha _{in}$,
identical phase conditions occur after a thickness increase of $\Delta D= 
\frac{2\pi }{k_z}=12.35$~\AA , thus leading to an accurate calibration of
the deposition rate.

The post growth GIXRD measurements consist of radial scans across
the (200) and (220) peaks in the in--plane ${\rm Cu_3Au(001)}$ reciprocal lattice.
These measurements were taken scanning the photon energy in broad ranges under
a suitable $\theta $-$2\theta $ scattering geometry.  
The observation of diffraction peaks in radial
scans allows to determine the in-plane spacing $d$ through the Bragg
condition $2d\sin \theta =hc/E$.
Rocking curves of selected Bragg peaks, obtained by rotating the azimuthal
angle at fixed energy, were measured to get additional information on the
surface morphology.

 A coverage estimate obtained by XPS measurements was found consistent
with the reflectivity measurements within $\pm \ 20\%$. ED polar scans were
measured by rotating the electron analyser in the plane defined by the
surface normal and the beam axis, while keeping the grazing angle, the
surface azimuthal orientation with respect to the beam axis and the
polarization orientation fixed. We considered emission along the two main symmetry
direction $\langle 100\rangle $$_{sub}$ and $\langle 110\rangle $$_{sub}$ 
of the substrate unit cell. The photon energy was set
to about 900 eV in order to look at several photoelectron and Auger peaks of
Fe, Cu and Au. Here we will focus on the Fe L$_{2,3}$M$_{23}$M$_{45}$ line
at a K. E. of 698 eV. The signal was collected at the maximum and at
suitably chosen energies aside the peak, in order to allow an effective
subtraction of the background of secondaries.

\section{Results}

\label{sec:risultati}

\subsection{GIXRD}

\begin{figure}[tbp]
\includegraphics[width=.5\textwidth]{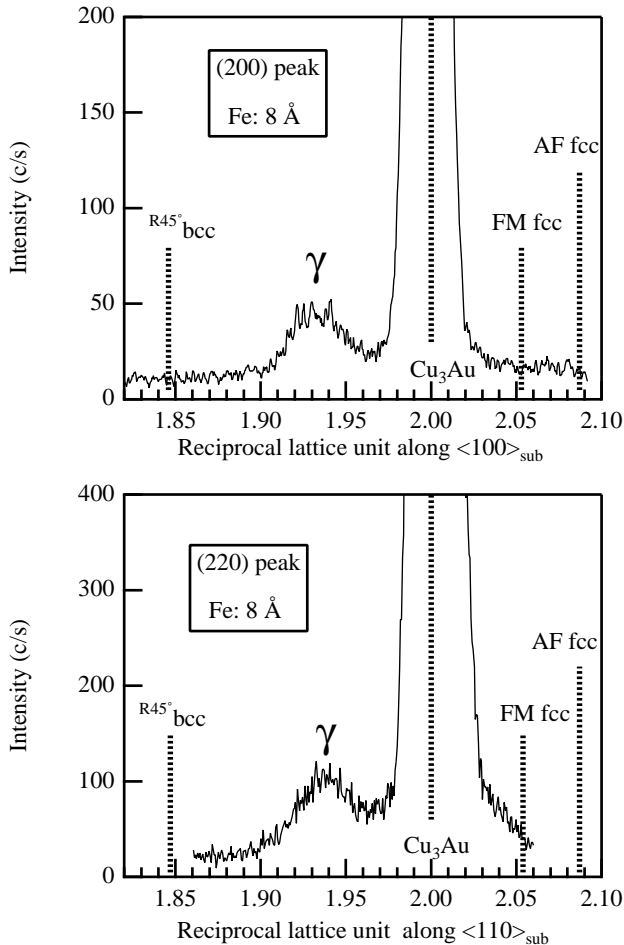}
\caption{In plane X-Ray Diffraction of the (2,0,0), upper panel, and 
(2,2,0), lower panel, diffraction peaks taken 
at fixed scattering geometry by varying the photon energy. The photon beam
impinges the surface at grazing incidence, forming an angle $\theta 
\sim 45^{\circ }$ with respect to the fcc(100) planes of the direct lattice.
The data are 
shown as a function of the Cu$_{3}$Au(100) reciprocal lattice unit in 
the corresponding lattice direction. 
Both diffraction patterns 
have been taken for the same Fe film at a thickness of $\sim$~8~\AA.
The vertical dotted lines correspond to the lattice of the 
antiferromagnetic (AF) and ferromagnetic (FM) fcc equilibrium 
structures as in Refs. [4,12]; the bcc(100) lattice, rotated 
by 45$^{\circ}$ is also shown.}
\label{fig2new}
\end{figure}

{\it 
Radial scans of the substrate diffraction peaks, taken on films of 
different thickness, are expected to show additional diffraction 
features arising from the Fe overlayers, if the film unit cell, although 
strained, is oriented parallel to the substrate, as reported in the 
literature. Upon deposition at room temperature, the first Fe-induced 
peak arises at a thickness of $\sim$~8~$\AA$. The radial scans across 
the (200) and (220) substrate peaks are shown in Fig.~\ref{fig2new} as a 
function of the substrate reciprocal lattice unit. The peaks labelled 
$\gamma$~ mark the appearance of a new structure, i.e. the overcome 
of the limit for pseudomorphic growth. The corresponding lattices for the 
antiferromagnetic and ferromagnetic fcc Fe are also shown for 
comparison,\cite{jonamarcus,wuttigS} as well as the lattice of a 
bcc(100) unit cell rotated by 45$^{\circ}$. Azimuthal scans taken on 
the $\gamma$~ peaks (not shown) indicate this Fe phase to be 
oriented parallel to the substrate lattice.
}

\begin{figure}[tbp]
\includegraphics[width=.5\textwidth]{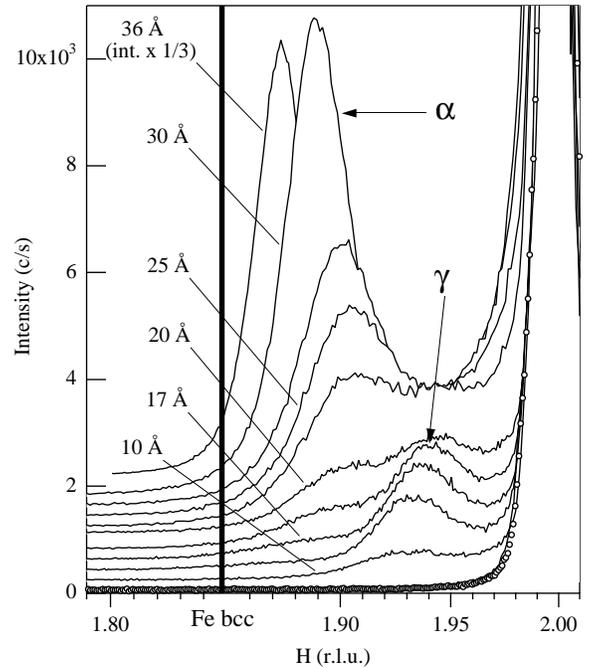}
\caption{In plane X-Ray Diffraction of the (2,0,0) diffraction peak taken 
at fixed scattering geometry by varying the photon energy. 
The data are shown as a function of the reciprocal lattice unit H of the $fcc$ 
Cu$_{3}$Au(100) substrate. Diffraction from the clean substrate is shown 
with open markers. Full lines are the diffraction measurements from 
Fe films at different thickness. The data have been vertically shifted by 
a constant offset (200~c/s) for the sake of clarity.}
\label{fig3new}
\end{figure}

The evolution of the Fe film structure as the thickness increases is 
shown in Fig.~\ref{fig3new}, where radial scans across the (200) 
substrate peak are shown from 3 up to 36~$\AA$.
The peak $\gamma$~ becomes more intense in the 10-17~\AA\ range.
A shoulder appears at the left hand side of peak $\gamma$ at 
$\Theta \geq$ 17~\AA . For $\Theta \geq $ 25~\AA \thinspace  this shoulder
develops in the well defined peak $\alpha $, that gradually moves away from
the (200) reflection.

The position $1.93\pm 0.005$~r.l.u. of the $\gamma$ peak in the 10~\AA~ 
pattern of fig.~\ref{fig3new}, is the same of Fig.~\ref{fig2new} at 
8~\AA~ and corresponds to a distance d$_{100}$ =$1.94~\pm
~0.01$~\AA\ along the $\langle 001\rangle $ direction of direct space,
eventually leading to a square cell of side 
$a_{{\alpha}'}$ =$d\cdot \sqrt{2}$ = 2.74~$
\pm $~0.01~\AA. 
The position of the $\alpha $ peak at 36~\AA\ thickness in Fig.~\ref{fig3new}
corresponds to a lattice parameter $a_{\alpha}$ = 2.830~$
\pm $~0.005~\AA . The position of $\alpha $ is indeed close
to the (220) reflection of a $^{R45}$bcc structure. Further, reasonable
arguments on the energetics of iron modifications and inspection of the
so--called epitaxial lines calculated in Ref.~\cite{jonamarcus} led us to
consider the $\alpha $ peak incompatible with an fcc-like 
modification (fct phase). 
The $\alpha $ peaks were therefore assigned to tetragonally strained body
centered ( $^{R45}$bct ) structures.

The GIXRD measurements provided therefore valuable insight on the in--plane
structure of the films; this information is relatively direct for those
patterns presenting one dominant Fe--induced peak, as it is the case at very
low ($\geq$ 8~\AA ) and very large coverages (36~\AA ). In an intermediate region
between 10 and 20~\AA , the patterns are more complex, likely reflecting the
coexistence of more than one phase in the film.
A full structural determination requires also the value of the vertical
lattice constant of the various iron phases which forms at different
thickness. We extracted this information by ED, in the forward focusing
regime.

\subsection{ED}

\begin{figure}[tbp]
\includegraphics[width=.5\textwidth]{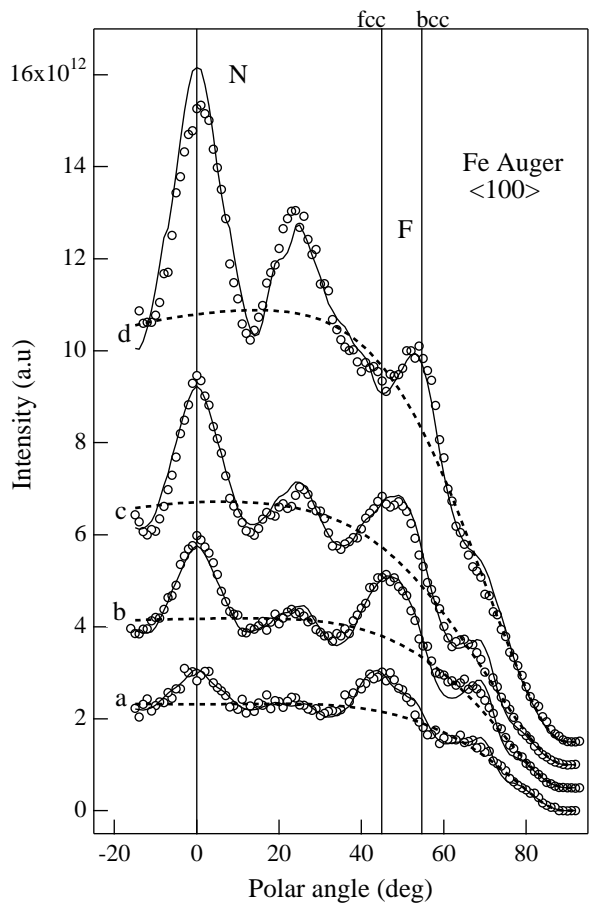}
\caption{Polar scans taken for the Fe Auger LMM (K.E. = 698 eV) peak along 
the $\left\langle100\right\rangle$ substrate symmetry direction for: a) the 3
\AA\ film, b) the 10 \AA\ film, c) the 20 \AA\ film, d) the 36 \AA\ film.
The polar scans are to scale and b), c), d) have been vertically 
shifted by a constant offset for the sake of clarity. 
Full lines are the fits to the experimental ED data (open circles). 
The heavy dashed lines represent the $ISO_{calc}$ components.
The surface normal direction ($N$) and the first neighbor direction ($F$) 
are also indicated by the vertical full lines. a) fit with a relaxed
pseudomorphic phase ($a_{p} = 2.65$~\AA, $c/a_p = 1.38$); b) fit with a model 
which combines the relaxed pseudomorphic phase and the $\gamma$ one
(see text for the values of the lattice 
parameters and the weight of the phases); 
c) fit with a linear combination of three phases (pseudomorphic, $\gamma$, and 
$\alpha$), see text for the values of the lattice 
parameters and the weight of the phases; 
d) fit with a body centered tetragonal model 
structure  ($a_{\alpha} = 2.83$~\AA~ and
 $c/a_{\alpha} = 1.03$).}
\label{fig4new}
\end{figure}

\begin{figure}[tbp]
\includegraphics[width=.5\textwidth]{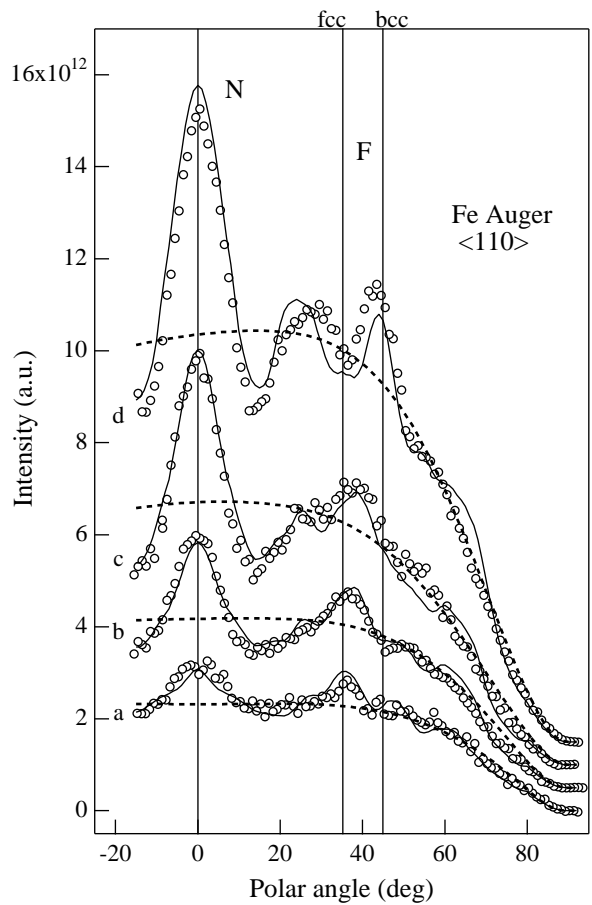}
\caption{Same of Fig.~4, but for the $
\left\langle110\right\rangle$ substrate symmetry direction for: a) The 3
\AA\ film, b) the 10~\AA\ film, c) the 20~\AA\ film, d) the 36~\AA\ film.
Full lines are the fits to the experimental ED data (open circles). The heavy 
dashed lines represent the $ISO_{calc}$ component, which is the same 
for both azimuthal directions at the corresponding Fe thickness.}
\label{fig5new}
\end{figure}

ED polar scan measured on films of selected thickness for the Fe Auger line along 
the $\langle 100\rangle _{sub}$ and $\langle 110\rangle_{sub}$ directions, 
are reported in Figs.~\ref{fig4new} and ~\ref{fig5new}, respectively.
The Fe films are the same of Fig.~\ref{fig3new} at the corresponding 
thickness.  
The patterns were obtained after subtraction of the background of secondaries.  
The peak F is originated by the forward scattering effect along off--normal
nearest--neighbour chains.  
The position of this peak provides a guess on the ratio between the in--plane and 
vertical lattice constants of the film.  
In this respect, the full vertical lines in Figs.~\ref{fig4new} 
and \ref{fig5new} mark the position of the F peak expected in case
 of fcc and $^{R45}$bcc geometry, respectively.  

At the lowest coverage investigated ( patterns $a$ of 
Figs.~\ref{fig4new} and \ref{fig5new}), 
the forward focusing peak F is fairly
pronounced in the $\langle 100\rangle _{sub}$ scan close to the fcc--like position . 
The F peak lays close to the fcc marker
in the $\langle 110\rangle _{sub}$ direction as well, though its intensity
is very weak. Note in both patterns the presence of the forward
focusing peak N, related to close--packed chains along the surface normal.
At 10~\AA~ (pattern $b$ of Figs.~\ref{fig4new} and \ref{fig5new})  the F peaks 
move slightly from the fcc markers towards larger polar angles. The angular shift
of the F peaks becomes more evident at $\Theta $~=~20~\AA\ (patterns $c$).
Also the shape of the patterns shows evident variations with respect to
lower coverages.
In the spectra obtained on the film at $\Theta $~=~36~\AA~ (patterns $d$), the F
peak becomes narrower and shifts towards the $bcc$ position.

We performed also ED measurements for the Auger lines of Cu and for the  Au
4f$_{7/2}$ photoemission line. The experimental data resulted in overall
agreement with results of Ref.~\cite{luches}, obtained on the same sample
and under similar experimental conditions of this paper. We address the
reader to Ref.~\cite{luches} for a careful discussion of the intermixing
between iron and substrate species.

\section{ED\ Data Analysis}
\label{sec:analisidati}

In this section we present the structural models for the Fe film at
different thickness, as obtained by comparison of the ED experimental data
to multiple scattering calculations.
In our computational approach the ED polar pattern $I_{exp}(\theta )$ is
considered as a superposition of two contributions, according to the
expression:

\begin{equation}
I_{exp}(\theta)=ISO_{exp}(\theta)\cdot(1+\chi_{exp}(\theta))
\end{equation}

where $ISO_{exp}(\theta )$ is a smooth, nearly isotropic background 
and the anisotropy term $\chi _{exp(\theta )}$ is the diffractive
part of the pattern, carrying the information on the interatomic distances.

The $ISO_{exp}$ contribution is commonly obtained by interpolating $
I_{exp}(\theta )$ to a polinomial and divided out in order to extract $\chi
_{exp}$, which is then compared to calculated $\chi _{calc(\theta )}$\cite
{gazzadi}. We have preferred to afford a calculation of the ISO term. Thus, the polar scans $
I_{exp}(\theta )$ have been compared with calculations:

\begin{equation}
I_{calc}(\theta )=ISO_{calc}(\theta )\cdot (1+\chi _{calc}(\theta ))
\end{equation}

For any given model the anisotropy term $\chi _{calc}$ has been calculated
by means of the MSCD code package \cite{MSCD}, while the ISO term has been
written as the product of several factors.\cite{fadiso,phdbruno}
First, an emission factor accounts for the electron emission matrix element;\cite{cooper}
it is determined by the polarization of the beam and by the initial and
final states of the emitted electron. A second factor bears the information
on the excited electron escape path through the Fe film;\cite{fadiso} it depends on the
thickness and the homogeneity of the film. 
A third factor accounts for the surface roughness.\cite{russo} Finally,
an instrumental factor accounts for the beam spot size on the sample and the
angular acceptance of the detector. The calculation of the ISO part requires
therefore reasonable estimates on the thickness and the surface roughness of the film
and on the electronic mean free path.

Concerning MSCD calculations, the non structural input parameters, i.e.
Multiple Scattering order~=~6 and the inner
potential~=~10~eV have been fixed for all the simulations. Clusters of at
least 180 atoms have been considered. The isotropic
emission for the Auger electrons has been simulated by the transition from
an initial p--level to an s ({\em l-1)} final--state level.

The value of the lattice parameter of the in--plane square cell $a$ has been
obtained from the GIXRD measurement, while the ratio between the vertical
spacing $c$ and $a$ , was first estimated by visual inspection of the
angular position of the F peaks in the ED patterns. With these input
parameters, the structural models have been refined by calculating $\chi
_{calc(\theta )}$ as a function of $a$ and $c$.

\subsection{Low Coverage: $\Theta < 8$ \AA}

The patterns of Figs.~\ref{fig4new} and \ref{fig5new} force us to consider an
fcc--like structure. Although results from Ref.~\cite{luches,ass} suggested
a limited degree of mixing of Au in the first few layers, a reasonable
fitting of the data was possible disregarding atomic exchange processes.\cite
{temperature} The comparison between MSCD calculations and the experimental
ED scans  is reported in Figs.~\ref{fig4new},\ref{fig5new} (curves 
labelled $a$). Note that the same 
$ISO_{calc}(\theta )$ is used in both azimuthal directions. 

The polar scans compare rather well with a simple
three-layer Fe pseudomorphic fcc film ( $a$= 2.65 ~\AA ; $c/a$ ~= ~1.41) 
built on an unrelaxed
substrate extending three layers beneath. 
However, a slightly better agreement )
was found by admitting a slight tetragonal distortion with $a_p $ =
2.65~\AA\ and $c/a_p =$~1.38.
{\it The quality of the fit is observed to be slightly worst along the 
$\langle 110\rangle _{sub}$ direction. In fact, the ED polar scans are 
certainly affected by the film morphology\cite{EDmorpho} (such as preferred step 
orientation), but we cannot exclude a structural origin related to a 
very slight zig-zag of the surface atom chains, which would smear the forward 
focussing features. In fact, a strong buckling was reported for the Fe on Cu(001) 
system,\cite{muller} where this distortion is predicted to be precursor of the 
fcc(100) to bcc(110) martensitic transition, \cite{spisak} however 
the latter transition is not observed on the present system.}

\subsection{Medium Coverage: 8 \AA\ $\leq \Theta \leq $ 20 \AA}

The $\gamma$ peak in the GIXRD patterns at 
$\Theta \geq 8$~\AA~ indicates a change of the growth mode. 
We first calculated the simulation for a homogeneous iron
phase made of six complete layers, assuming the $c/a$ ratio of 1.32
estimated by simple inspection of the angular position of the maximum of the
F peak in the polar pattern. This model yields a simulation of the ED data 
(not shown) definitely not adequate.
GIXRD and ED evidences can be rationalized if the
coexistence of the two phases is assumed: the majority fcc--like phase, as
seen at low coverage, is accompanied by the nucleation of a new phase 
$\gamma$, minority phase at this stage.

The ED patterns for the 10~\AA~ Fe film thickness (curve $b$ in 
Figs.\ref{fig4new},\ref{fig5new}) have been compared to calculations 
performed on a crude
model assuming the linear combination of two ''independent'' phases:
\begin{equation}
I_{th} = A\cdot I_p + B\cdot I_{\gamma} ,
\end{equation}
where

I$_p \rightarrow a_p $ = 2.65 \AA ~ and $c/a_p \sim 1.38$

I$_{\gamma} \rightarrow a_{\gamma}$ = 2.74 \AA 
~( from GIXRD) and $c/a_{\gamma}$ to be determined.

$c/a_{\gamma},$ $A$ and $B$ were varied in order to find the
best agreement with experimental data. Taking into account the estimated
thickness of the film, we have modelled a film of about 5-6 layers. A good
agreement was obtained with $c/a_{\gamma}=1.22$, $A$ = 0.6 and $B$
= 0.4 for both azimuthal directions (full line in polar scan $b$ of 
Figs.~\ref{fig4new},\ref{fig5new}).

For comparison, we also calculated the simulation for a homogeneous iron
phase made of six complete layers, assuming the $c/a$ ratio of 1.32
estimated by simple inspection of the angular position of the maximum of the
F peak in the polar pattern. This model yields a simulation of the ED data 
(not shown) definitely worse than the mixed phases model.

At 20~\AA\ the F peaks are clearly shifted, indicating a change of the $c/a$
ratio. Backed by GIXRD, a model considering the superposition of three
phases was attempted:

\begin{equation}
I_{th}=A\cdot I_p +B\cdot I_{\gamma}+C\cdot I_\alpha
\end{equation}

For simplicity, the parameters for the pseudomorphic and the $\gamma$ phases 
were fixed at the  same  values found at lower coverage:

I$_p \rightarrow  a_p $=2.65~\AA , $c/a_p \sim $ 1.38

I$_{\gamma} \rightarrow a_{\gamma}$ = 2.74 ~\AA , $c/a_{\gamma}$ = 1.22. 

By taking A = 0.4, B = 0.4 and C = 0.2, a rather satisfactory 
agreement was found (curves 
labelled $c$ in Figs.~\ref{fig4new},\ref{fig5new}) with the following set of
parameters for the $\alpha$ phase,

I$_{\alpha} \rightarrow a_\alpha $ = 2.80~\AA, $c/a_\alpha =1.05$.

We note that a reasonable
agreement was found also with a linear combination of the 
pseudomorphic phase ($a_p $=2.65~\AA, 
$c/a_p \sim 1.38$) and the $\gamma$ phase ($a_{\gamma}$= 2.74~\AA , $c/a_{\gamma}=1.22$) 
found at 10~\AA.  However, in this case, we have found 
rather different values of the $A$ and
$B$ coefficients along the two azimuthal directions
considered. Finally, the simulation for a homogeneous phase model assuming $c/a=1.25$, 
corresponding to the angular position of the F peak in the polar scan,
gave a bad quality fit.

\subsection{ High Coverage: $\Theta >$ 25 \AA }

The in--plane lattice parameter of the $\alpha $ phase at 36~\AA\ determined by GIXRD
is $a_\alpha = 2.830\pm 0.005$~\AA.  The value of
$c/a_\alpha $ was determined by simulation of scattering from a free standing 
film consisting of a unique phase of 10 layers.  The best fit
yielded $a_\alpha =2.83$~\AA~ and  $c/a_\alpha =1.03\pm 0.02$.  
The best fit curves along $\langle 100\rangle _{sub}$ and $\langle
110\rangle _{sub}$ are reported in Figs.~\ref{fig4new},\ref{fig5new} (curves 
labelled $d$). 
We found an excellent agreeement along the $\langle 100\rangle _{sub}$ 
direction. The simulation reproduces rather accurately the angular position,
intensity and width of the N and F peaks. The
experimental data are not  reproduced equally well along the 
$\langle 110\rangle _{sub}$ direction, possibly due to some morphological
effect.\cite {EDmorpho} A similar analysis, performed on the film of 
30~\AA\ thickness provided $a_\alpha =2.810\pm 0.005$~\AA~ and 
$c/a_\alpha =1.05\pm 0.02$.

\section{Discussion}
\label{sec:discussione}

\begin{figure}[tbp]
\includegraphics[width=.5\textwidth]{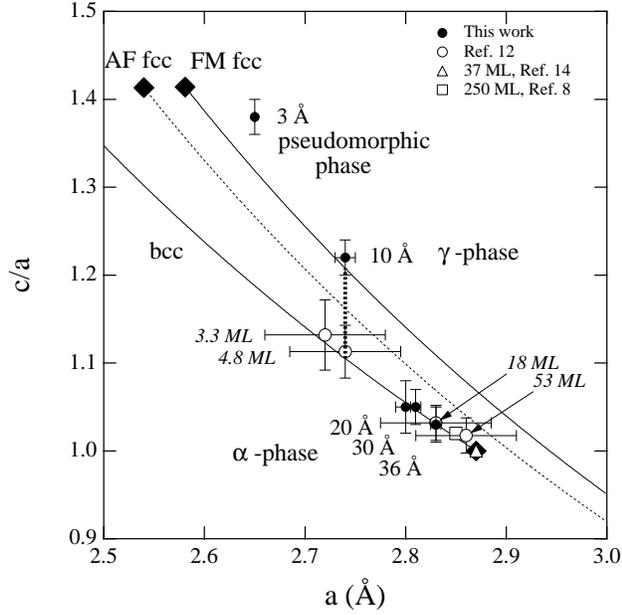}
\caption{Comparison of our results and
other experimental data available in literature for thick films  with the
theoretical prediction for the structure of the bct [after Ref.~4] and 
fct [after Ref.~12] phases, the so called
epitaxial curves (full and dashed lines). The full circles represent our 
structural determinations, the thickness and corresponding Fe phase 
are indicated for each point. The
open circles represent the determinations of Ref.~[12], the 
corresponding thickness is also indicated (in italics). 
The open triangle and the open square represent the measurement of Ref. 
[14] at 37~ML and that of Ref.~[8] for a 250~ML film,
respectively. The path for the transition from the fct to bct phase 
is indicated by the heavy dotted line.}
\label{fig6new}
\end{figure}

In Fig.~\ref{fig6new}, we report  the comparison of our results and
the other experimental data available in literature with the
theoretical prediction for the structure of the bct and fct phases, the so called
epitaxial curves\cite{jonamarcus} (i.e. the curves where the Poisson 
ratio between the elastic constants for the given structure is 
conserved).

The single phase we found at the lower coverage represents the pseudomorphic
phase predicted at initial stages of metal-on-metal epitaxy.\cite{vandermerwe} 
The occurrence of a pseudomorphic phase at low coverage is also
reported by other measurements on Cu$_3$Au(001) \cite{kirschnerM,luches} 
and Cu$_{90}$Au$_{10}$(001).\cite{lastkir}
As can be seen, this phase lies above the FM fcc 
epitaxial line and is locked by the substrate for a few layers 
before the appearance of the $\gamma$-phase 
at  $\sim 8$~\AA~
(for the pseudomorphic phase this thickness corresponds to 
4.5~ML), which indicates that the limit for pseudomorphic growth 
has been overcome (in Fig.~\ref{fig6new}, we have reported the point at 
10~\AA~ for which the vertical spacing has been also determined by 
the ED analysis). 
These data are therefore consistent with the limit of 4 ML 
found in a recent experiment \cite{luches} and
with the values obtained in earlier studies on this system
\cite{earlyjona,kisker1}.

A relevant exception to such general agreement on initial stages of growth,
is provided by the accurate analysis of Wuttig and coworkers.\cite{wuttigM,wuttigS} 
As can be seen in Fig.~\ref{fig6new}, the films were found to lie on the 
bcc epitaxial line already at the coverage of 3.3~ML (i.e. 
$\sim$~5~\AA, according to the data in 
table I of Ref.~\cite{wuttigS}). 
This discrepancy could be possibly attributed to the different procedure 
for film deposition, since the authors of Ref.~\cite{wuttigS} deposited the first 
two monolayers at 150 K, while subsequent layers were deposited at 300 K. 
This procedure is expected to reduce both intermixing and surface 
segregation. In fact, a fraction of the order of
10\% of a monolayer of Au atoms has been found to segregate 
for $\Theta$~$<$~4~ML upon deposition at RT\cite{luches}
and a few percents has been found within the Fe 
film\cite{kisker1,wuttigS,luches}. We may speculate that Au and Cu 
impurities are concurrent in the stabilization of the pseudomorphic 
phase. 
It is worth noting that a bcc-like structure has been recently 
suggested for 2-4 ML Fe films on Cu(001), where segregation 
is certainly much lower \cite{biedermann}.

The morphology of the growing film is also affected by 
the deposition procedure.
In fact, we previously found by He atom scattering that deposition of 
1 layer equivalent at 150~K and subsequent annealing at 400~K yields the 
formation of an homogeneous pattern of three-layer height 
islands,\cite{canepa} whereas a much higher filling of the first 
layer (68~\%) is obtained upon deposition of 1~ML at RT.\cite{Lin} 

Whatever the effect of temperature on segregation, structure and 
morphology, our measurements do not support a direct relation between 
the SRO transition and the fct to bct one. In fact, the latter phase 
only appears at a thickness of 17~\AA, i.e. well beyond the 
2.5-3.5~ML range claimed in the literature for 
the SRO transition.\cite{kisker1,wuttigM,kirschnerM}
On the basis of the present data, we cannot exclude a connection 
between the transition from the pseudomorphic phase to the fct 
$\gamma$-phase and the SRO transition.

The film evolution in the 10-20~\AA\ coverage range is the most 
interesting one since up to three different phases are seen to co-exist. 
The  $\gamma$ phase is seen to fall on the epitaxial 
line for FM fcc Fe.
Its lattice parameter at 8-10~\AA, $a_{\gamma}=2.74\pm 0.01$~\AA, 
seems to be the maximum allowed strain for this fct phase; in 
fact, when the third phase $\alpha$ appears at higher coverage,  
$a_{\gamma}$ rather shrinks (see Fig.~\ref{fig3new}). This suggests that the 
appearance of the $\alpha$ phase partially relieves the strain of the 
fct film. 
The $\alpha$ phase, when appearing at 17~\AA, has already a lateral 
lattice spacing of 2.80~\AA, which, together with the vertical spacing 
determined by MSCD analysis, brings the new phase directly on the bcc 
epitaxial line. No diffraction features are observed for lateral 
lattice spacings ranging from 2.74~\AA~ up to 2.80~\AA. 
Upon further deposition, the ''asymptotic'' orthomorphic $\alpha $ 
phase evolves along the bcc epitaxial line. 
The structure of this phase is seen to be fully consistent with 
previous data reported in the literature for thick 
films.\cite{kisker1,wuttigS,luches,kisker1}
From our data, the transition path between the fct and bct structure 
can be traced at the value of maximum tensile strain $a_{\gamma}=2.74$ 
of the $\gamma$ phase. The observation of a bct 
phase with the same lateral lattice constant was not possible, in fact 
the appearance of the $\alpha$ phase is accompanied by a partial 
relief of strain as witnessed by the slight decrease of $a_{\gamma}$, 
{\it which corresponds to a volume decrease of the $\gamma$ unit cell 
towards the predicted Fe fcc equilibrium point.}
 On the other hand, the limit value for the 
bct stability was appearently reached by the group of 
Wuttig.\cite{wuttigS} By following a different preparation procedure, 
they stabilized the bct phase down to four monolayers, with a 
corresponding lateral lattice constant in excellent agreement with the 
2.74~\AA~ value, we found for the limit of stability of the fct phase 
(see Fig.~\ref{fig6new}).

It is interesting to compare this system with the Fe/Cu(100) 
one. In the latter case,  Fe is also 
seen to grow with an fcc pseudomorphic structure for at least 4~ML, at 
higher coverage the magnetic properties change and the formation of  
buffer layers with anti-ferromagnetic fcc structure is proposed. 
An fcc(100) to bcc(110) 
transition takes place at about 10~ML, (which is also accompanied by 
an SRO transition) but, in this case, the whole film is 
observed to deconstruct, i.e. a martensitic phase transition takes 
place.\cite{martensitic,varga} 
This is not observed for Fe on Cu$_3$Au(001), where the $\gamma$ phase 
is still detectable at a thickness of 20~\AA.
In any case the formation of the bct phase is accompanied by a 
significant amount of surface roughness, possibly leading to the exposure of the 
layers closer to the Fe/substrate interface (which are still probed 
by ED at 20~\AA, see section III.B).
This is consistent with a quantitative evaluation of the bct domain size 
obtained by the width of the diffracted $\alpha$ peaks.  
The profiles of the $^{R45}$ bct (110) and $^{R45}$bct (100) peaks,
obtained with an azimuthal scan at fixed photon energy (not shown), yielded
a mean domain size of $\sim$~150~\AA~ for the 36~\AA~ film thickness.
This high level of morphological disorder is consistent with the 
experimental findings for
homoepitaxial deposition of Fe on Fe(001) at RT, where, in absence of surfactant
species, the growth is seen to proceed in a three-dimensional fashion,\cite{stroscio,bonanno}
resulting in the formation of pyramidal mound-like structures.\cite{Rieder}

\section{Conclusions}
\label{sec:conclusioni}

We have presented a combined GIXRD-ED study of the growth of thin films of
Fe on  Cu$_3$Au(001). We observed the formation of different 
phases as a function of the film thickness.
The main conclusions of this investigation can be
summarized as follows:\newline

i. A single pseudomorphic phase of nearly fcc character  was
observed below 8~\AA~ thickness. This
conclusion is consistent with most of the previous works in 
the literature\cite{earlyjona,kisker1,kirschnerM,luches} 
whereas it seems in apparent contrast
with the analysis of Ref.~\cite{wuttigS} that reports a bct phase down to 
$\sim 5$~\AA. In this respect, we suggested that exchange processes at the
interface \cite{luches} may concur to the stabilization of the 
pseudomorphic
phase and we speculate that the different results of Ref.~\cite{wuttigS} are
due to a lower influence of intermixing obtained during the deposition of
the first two layers at 150~K.

ii. On a 8~\AA~ Fe thickness film, a neat Fe induced peak,  
clearly illustrating the overcoming 
of the thickness limit for pseudomorphic growth, 
has been observed from the in--plane
GIXRD measurements.
Above this limit the film is characterized by the
coexistence of phases: the pseudomorphic phase, most
likely in inner layers, and a second strained phase $\gamma$. 
The latter phase is seen to lie on the ferromagnetic fcc epitaxial line. 

iii. At $\theta \sim$ 17~\AA, we clearly observed the nucleation of a 
third strained phase, $\alpha $, which becomes the dominating structure
at $\theta \geq$ 25~\AA. The $\alpha $ phase is characterized by a
body centered,  tetragonally strained structure,  whose unit cell is
rotated of 45$^{\circ }$ with respect to the fcc substrate. The strain is progressively
relieved with the increasing film thickness, in close agreement with the
results of Ref.~\cite{wuttigS}. The
experimentally determined lattice constants of $\alpha $ are in strict quantitative
agreement with the epitaxial curves for bcc Fe.

\section*{Acknowledgements}

Maurizio Canepa, Silvana Terreni and Lorenzo Mattera are very grateful
to Fernando Tommasini for his continuous support to this
experiment. The authors are then grateful to P. Cantini, G. Boato, S. Valeri
and A. Di Bona for stimulating discussions on the Fe/Cu$_3$Au system.
The authors thank G. Gazzadi for discussions on the MSCD calculations 
and C. Mannori and S. Prandi for helpful assistance at various stages of the
experiment.

Funding from INFM and from the Italian Ministero dell'Universit\`{a} e
Ricerca Scientifica ( Cofin 990211848) are gratefully acknowledged.

\newpage

\begin{references}
\bibitem{temperature}  The substrate temperature proved to be a very
critical parameter in determining the level of Au surface segregation. In
fact, the higher Au concentration at the Fe film surface, we found in the
preliminary experiments [Ref. 20], was associated with a slightly higher
(15 -20~$^{\circ}$C) substrate temperature during deposition.
\end{references}
\end{document}